\documentclass[twocolumn,prl,aps,groupedaddress,showpacs,floatfix]{revtex4}

\usepackage[dvipdfm]{graphicx}

\begin {document}
\title {Persistent Supercurrent Atom Chip}

\author {T. Mukai$^{1,3}$, C. Hufnagel$^{1,3,4}$, A. Kasper$^{1,3}$, T. Meno$^{3,5}$, A. Tsukada$^{1}$, K. Semba$^{1}$, and F. Shimizu$^{1,2,3}$}

\affiliation { $^1$NTT Basic Research Laboratories, NTT Corporation, 3-1, Morinosato-Wakamiya, Atsugi, Kanagawa 243-0198, Japan \\
$^2$Institute for Laser Science, University of Electro-Communications, 1-5-1 Chofugaoka, Chofu, Tokyo 182-8585, Japan\\
$^3$CREST, Japan Science and Technology Agency, 4-1-8 Honcho, Kawaguchi, Saitama 332-0012, Japan\\
$^4$Atominstitut \"Osterreichischer Universit\"aten, TU-Wien, Stadionallee 2, A-1020 Vienna, Austria\\
$^5$NTT Advanced Technology Corporation, 3-1, Morinosato-Wakamiya, Atsugi, Kanagawa 243-0198, Japan}

\date {\today}

\begin{abstract}

Rubidium 87 atoms are trapped in an Ioffe-Pritchard potential generated with a persistent supercurrent that flows in a loop circuit patterned on a sapphire surface. The superconducting circuit is a closed loop made of 100 $\mu$m wide MBE-grown MgB$_2$ stripe carrying a supercurrent of 2.5~A. To control the supercurrent in the stripe, an on-chip thermal switch operated by a focused argon-ion laser is developed. The switch operates as an on/off switch of the supercurrent, or as a device to set the current to a specific value with the aid of an external magnetic field. The current can be set even without an external source if the change is in decreasing direction.

\end{abstract}

\maketitle
%
An atom chip is a device to manipulate atoms near a solid surface with inhomogeneous magnetic fields generated by current carrying conductors constructed on a surface. It has been used to trap and guide atoms \cite{Chip_Guiding, Chip_BeamSplitter, Chip_AtomFiber}, to create quantum degenerate states of atomic gases \cite{Chip_BEC_Tuebingen, Chip_BEC_MPI, Chip_BEC_Heidelberg, Chip_Fermi} and to make atom interferometers \cite{Chip_Interferometer_Boulder, Chip_Interferometer_MIT, Chip_Interferometer_Heidelberg}. A merit of using micro magnetic structures is a strong trapping confinement that is achievable with a relatively small current. The shorter the distance between the potential minimum and the conductor, the larger the potential gradient gets, and atoms are more easily trapped or guided in a single spatial mode. However, a shorter atom-surface distance causes a faster decoherence. It has been reported that, when atoms are trapped with normal current at room temperature, inhomogeneous structures of wires for current, technical noise of the current source, as well as Johnson and RF noise by thermal fluctuations cause a large disturbance in the vicinity of a solid surface \cite{Surface_Tuebingen, Thermal_Spin_Flip, Henkel99_Euro, Henkel99_APB, Thermal_Loss}. For these reasons an atom chip with a smaller disturbance from the surface is strongly desired for high-quality on-chip applications like guided atom interferometers or quantum information processing devices.

Recently, the trapping of Rb atoms on a chip by using a supercurrent instead of a normal current was reported \cite{Alex, Haroche}. Since the resistance of a superconductor is zero, we may expect the reduction of current noise. Furthermore, the cryogenic temperature will help to reduce thermal noise and improve the background pressure, which contributes to the reduction of loss of atoms from the trap. In former experiments, however, the superconducting circuit was connected to the external current source that was operated at room temperature. We report in this paper the first demonstration of a magnetic atom trap that is generated by a persistent current on a closed superconducting circuit. The removal of external power source will further reduce the thermal disturbance. In addition it should be noted that the persistent current loop is a macroscopic quantum system. Since the atom-chip trap can hold various kinds of macroscopic quantum system of atoms including Bose-Einstein condensates, our system provides an unique opportunity to study physics of coupled macroscopic quantum systems of widely different origin. To manipulate atoms with closed supercurrent circuits, it is necessary to control currents on the chip. For this purpose we developed a laser-driven thermal switch \cite{Testardi} to control the supercurrent.

%
\begin{figure}[htbp]
\begin{center}
\includegraphics[height=5.5cm]{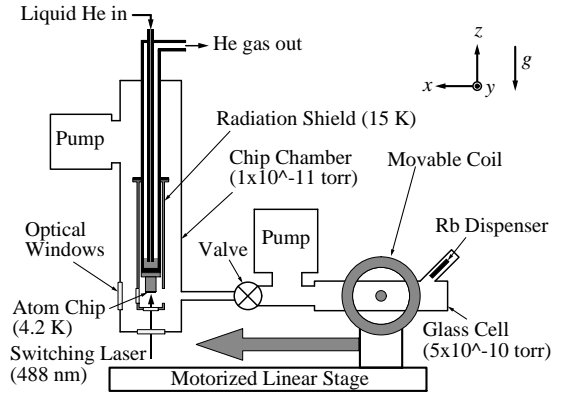}
\caption{The two chamber vacuum system. The atoms are first trapped and pre-cooled in the first chamber made of a glass cell. The cooling laser beams and coils for the bias magnetic field are not shown on this image. The  atoms are then loaded to the magnetic quadrupole trap and transported from the glass cell to the chip chamber by moving the quadrupole trap that is mounted on a motorized linear stage.}
\label{Vacuum}
\end{center}
\end{figure}

The experiment was done with a two chamber vacuum system (Fig.~\ref{Vacuum}) to keep small background pressure in the atom-chip region. The first chamber is a glass cell, where $^{87}$Rb atoms in the 5~$^{2}$S$_{1/2}$ $|F=1\rangle$ and $|F=2\rangle$ state are trapped in a conventional six beam magneto-optical trap (MOT) \cite{Raab} and cooled further by polarization-gradient cooling \cite{Dalibard,Ungar}. Atoms are then optically pumped to the $|F=2,m_{F}=+2\rangle$ state and are trapped in the quadrupole magnetic trap (QMT) that uses the same anti-Helmholtz coils as the MOT. The trapped atoms are transported to the second chamber by moving the coils for the magnetic trap. The first chamber is evacuated with an ion getter pump to $5 \times 10^{-10}$ torr. The second chamber is pumped with another ion getter pump together with a Ti-sublimation pump to $1 \times 10^{-11}$ torr. The two chambers are connected with a gate valve and a narrow tube ($\phi$ 10 $\times$ 60 mm) for differential pumping. The second chamber, which we call the chip-chamber, contains an atom-chip that is mounted on the bottom of a cold finger attached to a liquid Helium pot that can be cooled to 2.5~K by pumping helium vapor. The cryogenic system has a 2~W cooling power and the base temperature is usually kept at 4.2~K. The chip is shielded from black body radiation by a gold coated copper shield at 15~K. The radiation shield has five openings; one for the transfer of atoms and the other four that are equipped with AR coated glass windows ($\phi$ 20~mm) for optical access.

After the atomic cloud is transferred to the chip-chamber, it is adiabatically shifted upwards by applying an additional magnetic field pointing in the vertical direction (z-axis). When the center of the QMT is within the capture range of the chip potential, the quadrupole and the z-bias field are turned off and a magnetic field parallel to the y-axis is gradually applied. Then the atomic cloud is recaptured in an Ioffe-Pritchard type trap generated by the y-bias magnetic field and the persistent supercurrent.

%
\begin{figure}[htbp]
\begin{center}
\includegraphics[height=6cm]{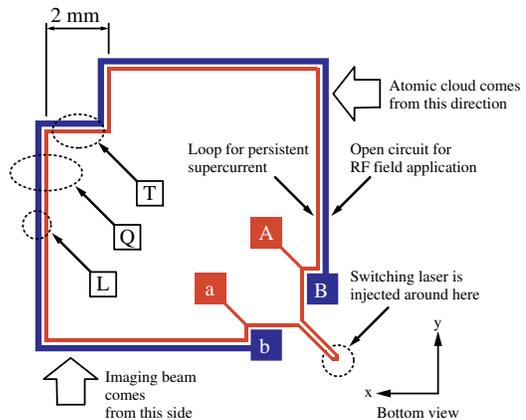}
\caption{(color online) The chip pattern of the persistent supercurrent atom-chip made from a 1.6 $\mu$m thick MBE-grown MgB$_{2}$ film. The inner loop (A-a: width = 100~$\mu$m) is used for a persistent current driving. The outer loop (B-b: width = 200~$\mu$m) is designed to apply radio frequency for evaporative cooling. Atoms are moved to the area marked by T for trapping and to Q for the distorted QMT measurement. The laser is illuminated on the appendix of the inner loop, or the area marked by L, when the dynamics of the thermal switch is studied.}
\label{PSChip}
\end{center}
\end{figure}

The persistent supercurrent chip was made from a 1.6~$\mu$m-thick magnesium diboride (MgB$_{2}$) film \cite{Ueda} that was grown by molecular-beam epitaxy (MBE) on a sapphire C substrate (10~mm $\times$ 10~mm $\times$ 0.5~mm). The transition temperature of MgB$_{2}$ is $T_{c}\approx$~35~K and the critical current density at 4.2~K is about $J_{c} \approx$~10$^{7}$~A/cm$^{2}$. The top of the MgB$_{2}$ thin film was coated with a thin gold layer to prevent radiation heating. The circuit pattern on the chip was produced by removing unnecessary part of the MgB$_2$ thin film by ion milling. The chip pattern is shown in Fig.~\ref{PSChip}. The circuit consists of two square loops. The inner loop is 100~$\mu$m wide and forms a closed circuit. It has a Z-shaped part at a corner for trapping atoms and an appendix at the diagonal corner which is used for the thermal switching. The outer open loop is 200~$\mu$m wide and was made to apply radio frequency for evaporative cooling. The current running through the Z-shaped part and a uniform bias magnetic field along the y-direction form an Ioffe-Pritchard magnetic potential. The total area enclosed by the inner loop circuit is $S = 6.80 \times 10^{-1}$~cm$^{2}$.

The temperature of the atom chip was kept at 4.2~K during the experiment. The persistent supercurrent, $I_{p}$, was induced into the loop circuit by the following method. Firstly, an argon-ion laser was focused on the tip of the appendix to heat it locally and to open the superconducting loop. Secondly, a magnetic field, $B_{\bot}$, perpendicular to the loop circuit was applied. Then, the argon-ion laser was turned off and the superconducting loop was closed. Finally, $B_{\bot}$ was turned off and the persistent supercurrent started running through the loop circuit. The magnitude of the supercurrent $I_{p}$ was controlled by $B_{\bot}$.

\begin{figure}[htbp]
\begin{center}
\includegraphics[width=7cm, bb=0 0 794 709]{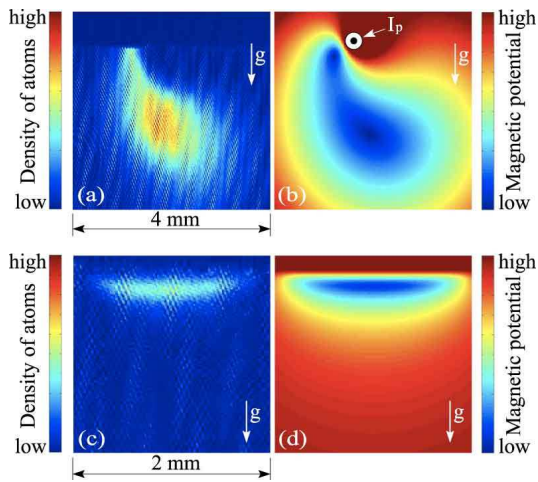}
\caption{(color online) (a) Pattern and density of the atoms in the QMT that is perturbed by the magnetic field induced by the persistent supercurrent on the chip (the field of view is 4 $\times$ 4~mm). (b) Calculated magnetic field potential with a 5.0 $\times$ 10$^{-3}$~T/cm quadrupole field and $I_{p}$ = 2.4~A. The position of $I_{p}$ is indicated with an arrow. (c) Pattern and density of the atoms in the chip trap (the field of view is 2 $\times$ 2 mm). The top edge of the figure is the position of the chip surface. (d) Calculated magnetic potential with $I_{p}$ = 2.5~A and $B_{y} = 1.5 \times 10^{-3}$~T.}
\label{Exp}
\end{center}
\end{figure}

The pattern and density of the atomic cloud were investigated by taking absorption images along the y-axis. Before we tried to trap atoms in the chip potential, we checked whether the persistent supercurrent had been correctly induced by the thermal switching. The atoms in the QMT were moved below the y-directed part of the loop (the ellipse marked by Q in Fig.~\ref{PSChip}) and an absorption image was taken along the y-axis (Fig.~\ref{Exp}(a)). In this image the field gradient of the QMT was 5.0 $\times$ 10$^{-3}$~T/cm, and the magnetic field that was used to induce the supercurrent was $B_{\bot} = 1.0 \times 10^{-3}$~T. The atomic cloud in the QMT shows clearly the distortion of the QMT potential caused by the persistent supercurrent on the chip. The potential shape of the magnetic trap generated by the QMT and the straight current on the chip was calculated at various current intensities. Figure~\ref{Exp}(b) shows equipotential curves which reproduce the atomic density pattern of the figure on the left. By comparing these images we estimated that the persistent supercurrent was approximately $I_{p}$ = 2.4~A.

To trap atoms in the persistent supercurrent chip potential, the atomic cloud was moved below the z-shaped part of the closed loop (the ellipse marked by T in Fig.~\ref{PSChip}) and transferred to the chip potential. Figure~\ref{Exp}(c) shows the image of atoms trapped in the chip potential after the QMT was switched off. Figure~\ref{Exp}(d) is the equipotential curves of magnetic field with $I_{p}$ = 2.5~A that fits best with the atomic pattern of the figure on the left. The small difference from the current estimated in Fig.~\ref{Exp}(b) is probably due to the magnetic field of the QMT that existed only at the measurement of the distorted QMT. With this current the magnetic flux trapped inside the loop was numerically evaluated to be $(6.6\pm0.2)\times10^{-8}$~Wb. This was in good agreement with the magnetic flux ($B_{\bot}S = (6.8\pm0.3)~\times~10^{-8}$~Wb) which was applied to induce the $I_{p}$. 
%
\begin{figure}[htbp]
\begin{center}
\includegraphics[height=5.5cm]{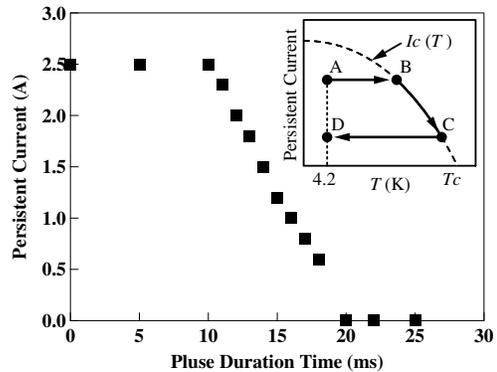}
\caption{Plot of persistent current $I_{p}$ left over after illuminating the switching laser pulse (0.7~W) as a function of the pulse duration time $\Delta \tau$. The inset shows the trajectory of  $I_{p}(T)$ in the switching process. During the initial rise of the temperature, $I_{p}$ is constant until it reaches $I_{c}(T)$ (the arrow from A to B). While $T$ rises further, $I_{p}$ adiabatically follows $I_{c}(T)$ (the arrow from B to C). When the laser is switched off, the persistent current $I_{p}$ is preserved and $T$ returns to the original value (the arrow from C to D).}
\label{CurrentLeft}
\end{center}
\end{figure}

To test the dynamics of the thermal switch we did the following experiments. When the laser is switched on, the temperature of the superconducting stripe starts to increase, and at some time the current $I_{p}$ flowing through the circuit becomes equal to the critical current $I_{c}(T)$. At this point the supercurrent should break. However, since the inductance of the circuit is small, $L\approx 27$~nH, the released power from the loop current is too small to completely break the superconductivity. Therefore, the supercurrent $I_{p}$ continues to flow keeping its magnitude to $I_{c}(T)$. (See the inset of Fig.~\ref{CurrentLeft}.) When the temperature reached $T_{c}$ the supercurrent becomes zero. This suggests that we can set $I_{p}$ to any value below the initial magnitude only by adjusting the duration time of the laser pulse. We illuminated the appendix of the inner loop by the laser and estimated the current left over in the loop from the image of trapped atoms. Figure~\ref{CurrentLeft} shows the $I_{p}$ persisted after the injection of the switching pulse (0.7~W) as a function of duration time $\Delta \tau$. The result shows that $I_{p}$ was kept constant in the initial 10~ms, decreased gradually during 10 $\le \Delta \tau \le$ 20~ms, and became zero at around 20~ms. In this experiment atoms were brought to the chip after the heating and cooling cycle was completed. They were trapped after the cycle, and $I_{p}$ was calibrated from the position and pattern of the trapped atoms by the same method used in the trapping experiment of Fig.~\ref{Exp}(c) and (d).
%
\begin{figure}[htbp]
\begin{center}
\includegraphics[height=5.5cm]{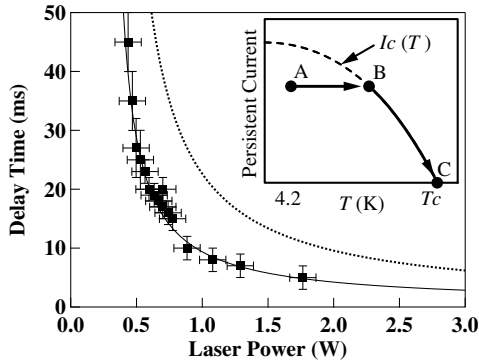}
\caption{Plot of the delay time $t_{d}$ from the onset of the laser illumination to the termination of persistent current as a function of the laser power $p$ deposited on the atom chip. The filled squares represent the experimental data when the switching laser is illuminated the appendix of the inner loop. The solid curve is the fitting by the empirical function, $t_d(p) = \frac{\alpha}{p-\beta}+\gamma$, with $\alpha = 6.5\times 10^{-3}$~J, $\beta = 0.27$~W, and $\gamma = 0.47\times 10^{-3}$~s. The dotted curve is the experimental result when the superconductivity was broken by illuminating the outer loop at L in Fig.~\ref{PSChip}. }
\label{Delaytime}
\end{center}
\end{figure}

We repeated the above experiment at various laser power and measured the time delay $t_{d}$ from the onset of the laser illumination to the termination of $I_{p}$ as a function of the laser power $p$.  The $t_{d}(p)$ fitted well to an empirical function
\begin{equation}
\label{SwitchDelay}
t_d(p)=\frac{\alpha}{p-\beta}+\gamma
\end{equation}
with the values $\alpha = 6.5 \times 10^{-3}$~J, $\beta = 0.27$~W, and $\gamma = 0.47 \times 10^{-3}$~s. By considering the limiting case, one may interpret that $\alpha$ is the minimum laser energy deposited in the chip to break the supercurrent, and $\beta$ is the minimum power of a continuous laser required to keep the heated part of the circuit above $T_c$. The constant $\gamma$ is caused by the delay of the mechanical shutter, which is used for the laser switching, from the timing signal.  Note that the minimum laser power required to switch the current is nearly one order of magnitude smaller than the cooling capacity of our cryogenic system, and that many switches can be installed on a chip. When the chip surface is heated, the heat conduction expands the heated area much beyond the size of the supercurrent stripe. This area is estimated from $\alpha$. Assuming the energy from the laser is deposited spherically, $\alpha \approx (\pi/3) c_v r^{3} \Delta T$, where $c_{v}$ is the heat capacity per volume of Al$_{2}$O$_{3}$ at $T_{c}$, $r$ is the radius of the sphere. In this expression we took into account that $c_{v}$ changes approximately proportional to $T^{3}$. Inserting $c_{v}(T_{c}) \approx 0.2$~JK$^{-1}$cm$^{-3}$ \cite{Heat_Capacity}, we obtain $r\approx 1$~mm. Therefore, if two thermal switches are to be installed they must be separated at least by 2~mm.

The thermal switch can be operated in principle at any point of the circuit. We measured also the switching characteristics of the outer loop by illuminating the laser at the point that is marked by L in Fig.~\ref{PSChip}. The termination of supercurrent was determined by the onset of a finite resistance between B and b terminals. The result is shown by the dotted curve in Fig.~\ref{Delaytime}. The parameter $\alpha= 15.4 \times 10^{-3}$~J is approximately 2.5 times larger than the case of the switching at the appendix. This, we believe, is due to the wider width of the outer loop conductor. The laser power required to switch should scale up with the width of the superconducting stripe.  Several other methods are possible to reduce the laser power requirement. Grooves on both sides of the conductive stripe reduces the thermal conductivity. Deposition of absorptive material on the stripe and a laser with longer wavelength will heat more efficiently and localize the heating of the chip.

In conclusion we have trapped atoms with a magnetic potential generated by a persistent supercurrent. The persistent supercurrent was controlled with an on-chip thermal switch driven by a laser. With a persistent supercurrent atom chip, a long decoherence time is expected even at a very small distance from the chip surface. Therefore, this technique will help to realize a spatial single-mode waveguide and trap for atoms, which is applicable for high-quality on-chip applications like guided atom interferometers or quantum information processing devices. In addition our system will provide an unique opportunity to study physics of coupled macroscopic quantum systems of widely different origin.

\section*{Acknowledgements}
The authors would like to thank T. Eichler of Intel Corporation for initial construction of the experimental setup, K. Ueda of NTT Basic Research Laboratories and M. Naito of Tokyo Univ. of Agriculture and Technology for the MBE-growth of MgB$_{2}$ thin film. We are also grateful to M. Ueda of Tokyo Institute of Technology for valuable discussions.

%

\bibliographystyle{prsty}


\end{document}